# Coupled exciton internal and center-of-mass motions in two-dimensional semiconductors by a periodic electrostatic potential

Fujia Lu[1], Qianying Hu[2,3], Yang Xu[2,3], Hongyi Yu[1,4]*

[1] Guangdong Provincial Key Laboratory of Quantum Metrology and Sensing & School of Physics and Astronomy, Sun Yat-Sen University (Zhuhai Campus), Zhuhai 519082, China

[2] Beijing National Laboratory for Condensed Matter Physics, Institute of Physics, Chinese Academy of Sciences, Beijing 100190, China.

[3] School of Physical Sciences, University of Chinese Academy of Sciences, Beijing 100049, China.

[4] State Key Laboratory of Optoelectronic Materials and Technologies, Sun Yat-Sen University (Guangzhou Campus), Guangzhou 510275, China

* E-mail: yuhy33@mail.sysu.edu.cn

**Abstract:** We theoretically investigated the coupling between the exciton internal and center-of-mass motions in monolayer transition metal dichalcogenides subjected to a periodic electrostatic potential. The coupling leads to the emergence of multiple absorption peaks in the exciton spectrum which are the hybridizations of 1$s$, 2$s$ and 2$p_\pm$ Rydberg states with different center-of-mass momentums. The energies and wave functions of hybrid states can be strongly modulated by varying the profile of the periodic electrostatic potential, which well reproduces the recent experimental observations. Combined with the electron-hole exchange interaction, non-degenerate valley-coherent bright excitons can be realized by applying an in-plane electric field, with the valley coherence determined by the field direction.

## I. Introduction

The exciton is a hydrogen-like bound state formed by an electron and a hole through their Coulomb interaction. In atomically thin layers of semiconducting transition metal dichalcogenides (TMDs), exceptionally strong Coulomb interactions are exhibited between charged carriers due to the reduced dielectric screening in two-dimensional (2D) systems. Therefore, excitons play a key role in photonic and optoelectronic properties of TMDs [1-3]. A free exciton can be viewed as a two-body system with a wave function consisting of a center-of-mass (CoM) part and an internal part describing the electron-hole relative motion. Similar to the 2D hydrogen atom, the exciton internal motion manifests as a series of discrete Rydberg states 1$s$, 2$s$, 2$p_\pm$, … [4-7]. For these free excitons, $s$-type Rydberg states with zero CoM momentums are optically active and feature large oscillator strengths. Excitons in monolayer TMDs are found to exhibit various novel properties, including strong binding energies and small Bohr radii [8-12], non-hydrogenic Rydberg series [4-6], valley-dependent optical selection rules [13-17] and tunable exciton valley pseudospin through optical and magnetic fields [18-24].

Lateral superlattice potentials in 2D materials can serve as a novel platform for exploring exotic quantum phenomena [3,25]. In van der Waals stackings of layered 2D materials, the formation of long-wavelength moiré patterns with spatially modulated atomic registries can naturally introduce a superlattice potential [26-29]. Recent experiments have revealed rich correlated insulating states [30-34] in moiré patterns of bilayer TMDs, demonstrating their great potential for quantum simulation. Besides, carriers in 2D layered materials are susceptible to external perturbations due to the atomically thin geometry, making possible the external implementation of superlattice potentials. In monolayers with the absence of moiré patterns, superlattice potentials with wavelengths ~ 100 nm have been introduced by periodic strain patterning [35,36]. Meanwhile, the spatially inhomogeneous charge distribution in moiré patterns of twisted TMDs, hexagonal boron nitride (hBN) and graphene layers can also generate periodic electrostatic potentials in adjacent 2D layers, with wavelengths ~ 10 nm adjustable through the twist angle [37-43]. In these moiré systems, twisted bilayer graphene (TBG) has attracted widespread attention for its large and dynamically tunable carrier density. The moiré patterned TBG can generate strong superlattice potentials in adjacent 2D layers and significantly alter their electronic properties.

Early studies about the moiré effect on excitons focus on the modulation of exciton CoM motion by the superlattice potential [28,29]. Very recently, several experiments have shown that the absorption of high-energy Rydberg excitons in monolayer TMDs can be largely affected by the carrier distribution in an adjacent TBG moiré pattern [40,43]. When the moiré wavelength reaches $\lambda$ ~ 20 nm or larger, monolayer TMDs exhibits multiple absorption peaks near the energy of 2s Rydberg exciton, which red-shift significantly when increasing the carrier density in TBG. In this work, we developed a theoretical model to describe the behavior of excitons in monolayer TMDs subjected to a periodic electrostatic potential. Our analysis indicates that the exciton internal and CoM motions are coupled by the periodic electrostatic potential, resulting in eigenstates being the hybridizations of s- and p-type Rydberg states with different CoM momenta. Oscillator strengths of 1s and 2s excitons are then redistributed into multiple hybrid states, giving rise to additional absorption peaks. The simulated exciton spectrum based on our model can well reproduce experimental observations. We also propose to realize non-degenerate valley-coherent hybrid excitons by applying an in-plane electric field, which have linearly polarized optical selection rules with the polarization direction determined by the field direction.

The remainder of this paper is organized as follows. In Sec. II we give theoretical models for excitons subjected to a periodic electrostatic potential. In Sec. III we present the calculated optical spectrum of excitons in the TMDs/TBG system under different doping densities and moiré wavelengths. Sec. IV discusses the tunable valley coherence of the exciton by an in-plane electric field. A brief conclusion is presented in Sec. V.

## II. Theoretical models for excitons under a periodic electrostatic potential

We consider a van der Waals heterostructure formed by a monolayer TMDs and an adjacent TBG moiré pattern as illustrated in Fig. 1(a). The spatially periodic charge density in TBG (see Fig. 1(b) for an illustration) can remotely generate a periodic electrostatic potential in monolayer TMDs. Writing the potential applied on the electron in monolayer TMDs as $U(\mathbf{r}_e)$, a hole then feels a potential $-U(\mathbf{r}_h)$. Here $\mathbf{r}_{e/h}$ is the spatial coordinate of the electron/hole. The exciton Hamiltonian in monolayer TMDs can be written as $\hat{H} = \hat{H}_X + \hat{U}_X$, with

$$\hat{H}_X = -\frac{\hbar^2}{2M}\frac{\partial^2}{\partial \mathbf{R}^2} - \frac{\hbar^2}{2\mu}\frac{\partial^2}{\partial \mathbf{r}^2} + V(\mathbf{r}),$$
$$\hat{U}_X = U(\mathbf{r}_e) - U(\mathbf{r}_h) = U\left(\mathbf{R} + \frac{m_h}{M}\mathbf{r}\right) - U\left(\mathbf{R} - \frac{m_e}{M}\mathbf{r}\right). \quad (1)$$

Here $\hat{H}_X$ corresponds to the free exciton Hamiltonian with $V(\mathbf{r})$ the electron-hole Coulomb interaction, and $\hat{U}_X$ is the total external electrostatic potential. $\mathbf{R} \equiv \frac{m_e}{M}\mathbf{r}_e + \frac{m_h}{M}\mathbf{r}_h$ and $\mathbf{r} \equiv \mathbf{r}_e - \mathbf{r}_h$ are the CoM and electron-hole relative coordinates, respectively. $M = m_e + m_h \approx m_0$ is the exciton mass ($m_0$ is the free electron mass), and $\mu = m_e m_h / M$ is the reduced mass. The eigenstate $|\mathbf{Q}, nl\rangle \equiv |\mathbf{Q}\rangle|nl\rangle$ of $\hat{H}_X$ can be separated into a CoM wave function $|\mathbf{Q}\rangle$ in the plane form, and an electron-hole relative wave function in the discrete Rydberg state $|nl\rangle$ ($nl = 1s, 2s, 2p_\pm, ...$) which describes the exciton internal motion [4-7]. The energy of $|\mathbf{Q}, nl\rangle$ is $\frac{\hbar^2 \mathbf{Q}^2}{2M} + E_{nl}$. Note that here we didn't take into account the electron-hole exchange interaction, which doesn't affect the main results and will be considered in detail in Sec. IV below.

$U(\mathbf{R})$ is a slowly varying potential in a length scale of the moiré wavelength $\lambda \sim 20$ nm, significantly larger than the exciton Bohr radius ($\sim 2$ nm for $1s$ state [10-12]). Consequently, we adopt a linear expansion $U\left(\mathbf{R} \pm \frac{m_{h/e}}{M}\mathbf{r}\right) \approx U(\mathbf{R}) \pm \frac{m_{h/e}}{M}\mathbf{r} \cdot \nabla U(\mathbf{R})$ and write

$$\hat{U}_X \approx \mathbf{r} \cdot \nabla U(\mathbf{R}) = i\mathbf{r} \cdot \sum_{\mathbf{G}\neq 0} \mathbf{G} U(\mathbf{G}) e^{i\mathbf{G}\cdot\mathbf{R}}. \quad (2)$$

In the above last step, we have expanded the periodic function $U(\mathbf{R})$ into the Fourier series $U(\mathbf{R}) = \sum_{\mathbf{G}} e^{i\mathbf{G}\cdot\mathbf{R}} U(\mathbf{G})$, with $\mathbf{G}$ the reciprocal lattice vector of the superlattice.

$\hat{U}_X$ couples two states $|\mathbf{Q}, nl\rangle$ and $|\mathbf{Q}', n'l'\rangle$ with an angular momentum difference $l - l' = \pm 1$ and a CoM momentum difference $\mathbf{Q} - \mathbf{Q}' = \mathbf{G}$, with the coupling matrix element

$$\langle \mathbf{Q}, nl | \hat{U}_X | \mathbf{Q}', n'l' \rangle = i \sum_{\mathbf{G}\neq 0} \delta_{\mathbf{Q}-\mathbf{Q}',\mathbf{G}} U(\mathbf{G}) \mathbf{G} \cdot \langle nl | \hat{\mathbf{r}} | n'l' \rangle$$
$$= \sum_{\mathbf{G}\neq 0} \delta_{\mathbf{Q}-\mathbf{Q}',\mathbf{G}} t^{nl}_{n'l'}(\mathbf{G}). \quad (3)$$

The exciton Hamiltonian can then be written in the form $\hat{H} = \sum_{\mathbf{Q}\in\text{mBZ}} \hat{H}_\mathbf{Q}$ (mBZ stands for the superlattice mini-Brillouin zone), with

$$\hat{H}_\mathbf{Q} = \sum_\mathbf{G} \sum_{nl} \left( \frac{\hbar^2(\mathbf{Q}+\mathbf{G})^2}{2M} + E_{nl} \right) |\mathbf{Q}+\mathbf{G}, nl\rangle\langle\mathbf{Q}+\mathbf{G}, nl|$$

$$+ \sum_{\mathbf{G}\neq\mathbf{G}'} \sum_{nl,n'l'} t^{nl}_{n'l'}(\mathbf{G}-\mathbf{G}')|\mathbf{Q}+\mathbf{G}, nl\rangle\langle\mathbf{Q}+\mathbf{G}', n'l'|. \quad (4)$$

By diagonalizing the above Hamiltonian, the $n_X$-th exciton branch becomes the hybridization of various Rydberg states:

$$|\Phi_{n_X,\mathbf{Q}}\rangle = \sum_\mathbf{G} \sum_{nl} \langle \mathbf{Q}+\mathbf{G}, nl|\Phi_{n_X,\mathbf{Q}}\rangle |\mathbf{Q}+\mathbf{G}, nl\rangle. \quad (5)$$

For optically active bright excitons, momentum conservation requires $\mathbf{Q} = 0$. Note that due to the mixing between different $|nl\rangle$, the angular momentum $l$ for the electron-hole relative motion is no longer a good quantum number. In experiments, $U(\mathbf{r}_{e/h})$ is generated by a triangular-type moiré pattern with the in-plane $2\pi/3$ rotation ($\hat{C}_3$) symmetry [37-43], see Fig. 1(b). In this case, $\hat{C}_3|\Phi_{n_X,\mathbf{Q}=0}\rangle = e^{-i\frac{2\pi}{3}C_{3X}}|\Phi_{n_X,\mathbf{Q}=0}\rangle$, where $C_{3X}$ corresponds to the sum of $\hat{C}_3$ quantum numbers for CoM and internal motions. Bright excitons must have $C_{3X} = 0$.

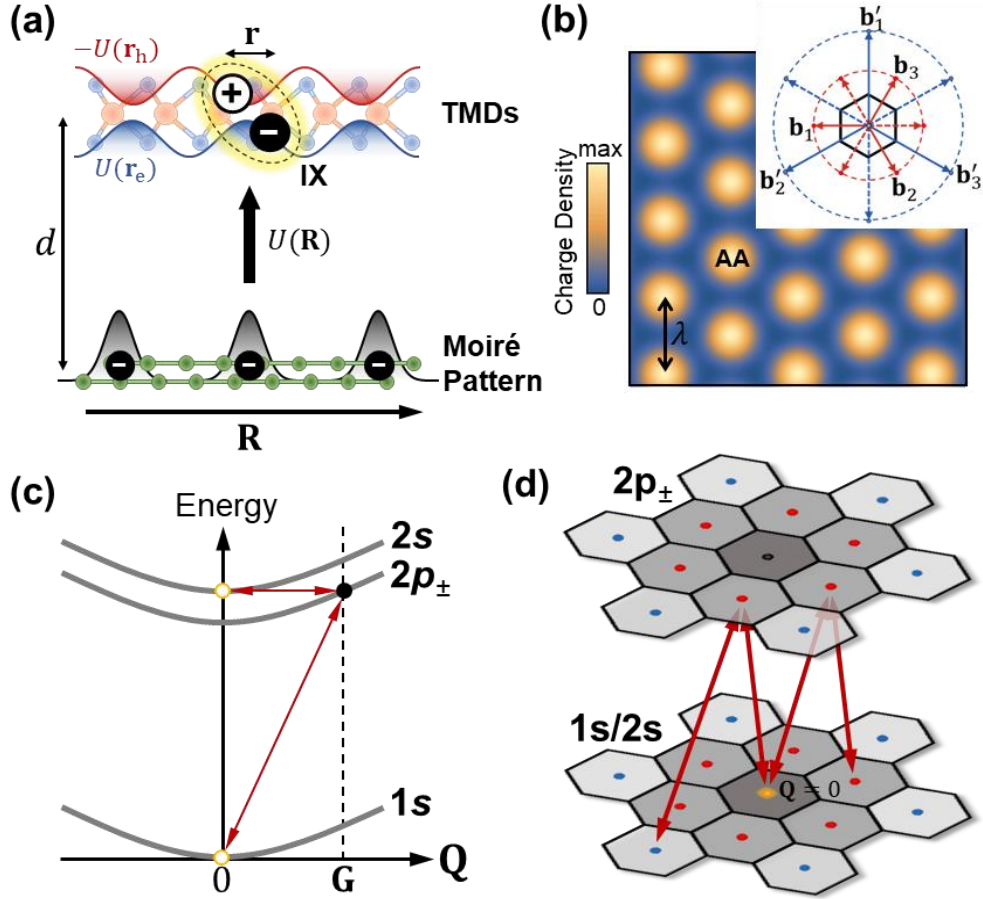

FIG. 1. (a) A schematic illustration of the van der Waals heterostructure formed by a monolayer TMDs and an adjacent TBG moiré pattern vertically separated by $d$. The periodic charge density in the moiré

pattern generates an electrostatic potential in the monolayer TMDs, where the electron and hole feel potentials $U(\mathbf{r}_e)$ and $-U(\mathbf{r}_h)$, respectively. (b) The 2D map of the charge distribution, with the density maxima located at **AA** sites in TBG. The inset shows the superlattice mini-Brillouin zone (black hexagon) and reciprocal lattice vectors $\pm \mathbf{b}_n$ (red arrows) and $\pm \mathbf{b}'_n$ (blue arrows). (c) Dispersions of free exciton Rydberg states $1s$, $2s$ and $2p_\pm$. $\mathbf{Q}$ is the exciton CoM momentum. Double arrows denote couplings induced by the periodic electrostatic potential. (d) A schematic of the oscillator strength redistribution. The $\mathbf{Q} = 0$ bright $1s$ and $2s$ excitons are coupled directly (indirectly) to dark $2p_\pm$ ($1s/2s$) excitons with finite CoM momentums, thus redistribute their oscillator strengths into multiple hybrid states.

Considering that $U(\mathbf{R})$ is a smooth function of $\mathbf{R}$, $|U(\mathbf{G})|$ should decrease rapidly with $|\mathbf{G}|$. Below we only consider $U(\mathbf{G})$ with $\mathbf{G} = \pm \mathbf{b}_{1,2,3}$ which are nonzero reciprocal lattice vectors with the smallest magnitude, see Fig. 1(b). $U(\mathbf{R})$ induced by a TBG moiré pattern is expected to be $\hat{C}_3$- and inversion-symmetric, resulting in $U(\pm \mathbf{b}_1) = U(\pm \mathbf{b}_2) = U(\pm \mathbf{b}_3) = U_0$. For the free exciton state $|\mathbf{G}, nl\rangle$, we shall only keep $\mathbf{G} = 0$, $\pm \mathbf{b}_{1,2,3}$ and $\pm \mathbf{b}'_{1,2,3}$, where $\pm \mathbf{b}'_1, \pm \mathbf{b}'_2, \pm \mathbf{b}'_3$ are nonzero reciprocal lattice vectors with the second-smallest magnitude, see Fig. 1(b). We also focus on the four lowest-energy Rydberg states with $nl = 1s$, $2s$ and $2p_\pm$, see Fig. 1(c). Because of the nonlocal screening effect of 2D layered materials, $E_{2s}$ is slightly higher than $E_{2p_\pm}$ [6]. Here we set $E_{2s} - E_{2p_\pm} = 5$ meV considering that the binding energies of $2s$ and $2p_\pm$ excitons are weak when near TBG [37,40]. The six degenerate states $|\pm \mathbf{G}\rangle$, $|\pm \hat{C}_3 \mathbf{G}\rangle$ and $|\pm \hat{C}_3^2 \mathbf{G}\rangle$ for the exciton CoM motion with $G \neq 0$ can be reformulated into a series of $\hat{C}_3$-symmetric basis states $|\psi_{C_{3,\text{CoM}},G}\rangle$ and $|\bar{\psi}_{C_{3,\text{CoM}},G}\rangle$ with even and odd parities, respectively, whose forms are summarized in Table I. Here $C_{3,\text{CoM}}$ represents the $\hat{C}_3$ quantum number of the CoM wave function, and the resultant total $\hat{C}_3$ quantum number of the exciton is $C_{3X} = C_{3,\text{CoM}} + l$ (mod 3), with $l = 0, +1$ and $-1$ for $s$, $p_+$ and $p_-$ Rydberg states, respectively. When referring to bright excitons with even parity and $C_{3X} = 0$, the involved basis states are $|\mathbf{Q} = 0, ns\rangle$, $|\psi_{0,b}, ns\rangle \equiv |\psi_{0,b}\rangle|ns\rangle$, $|\psi_{0,\sqrt{3}b}, ns\rangle$, $|\bar{\psi}_{-,b}, 2p_+\rangle$, $|\bar{\psi}_{+,b}, 2p_-\rangle$, $|\bar{\psi}_{-,\sqrt{3}b}, 2p_+\rangle$ and $|\bar{\psi}_{+,\sqrt{3}b}, 2p_-\rangle$.

After taking into account the coupling effect between exciton internal and CoM motions with a strength characterized by $|t_{ns}^{2p}| = U_0 b |\langle 2p_\pm | \hat{\mathbf{r}} | ns \rangle|$, a bright eigenstate $|\Phi_{n_X,\mathbf{Q}=0}\rangle$ becomes the hybridization of $|0, ns\rangle$, $|\psi_{0,b}, ns\rangle$, $|\psi_{0,\sqrt{3}b}, ns\rangle$, $|\bar{\psi}_{\mp,b}, 2p_\pm\rangle$ and $|\bar{\psi}_{\mp,\sqrt{3}b}, 2p_\pm\rangle$. The oscillator strengths of free excitons $|\mathbf{Q} = 0, 1s\rangle$ and $|\mathbf{Q} = 0, 2s\rangle$ are thus redistributed into a series of bright hybrid states with different energies, giving rise to multiple absorption peaks in the exciton spectrum as observed in experiments.

TABLE I. Equation forms of $\hat{C}_3$- and inversion-symmetric basis states $|\psi_{C_3,\text{CoM},G}\rangle$ and $|\bar{\psi}_{C_3,\text{CoM},G}\rangle$ for the exciton CoM motion with $\mathbf{G} \neq 0$. We have used the notations $\mathbf{G}_0 \equiv \mathbf{G}$, $\mathbf{G}_1 \equiv \hat{C}_3 \mathbf{G}$, $\mathbf{G}_2 \equiv \hat{C}_3^2 \mathbf{G}$ and $\bar{\mathbf{G}}_j \equiv -\mathbf{G}_j$.

| $C_{3,\text{CoM}}$ | **Even parity** for the CoM motion | **Odd parity** for the CoM motion |
|---|---|---|
| 0 | $\|\psi_{0,G}\rangle \equiv \frac{1}{\sqrt{6}} \sum_{j=0}^{2} (\|\mathbf{G}_j\rangle + \|\bar{\mathbf{G}}_j\rangle)$, | $\|\bar{\psi}_{0,G}\rangle \equiv \frac{1}{\sqrt{6}} \sum_{j=0}^{2} (\|\mathbf{G}_j\rangle - \|\bar{\mathbf{G}}_j\rangle)$, |
| +1 | $\|\psi_{+,G}\rangle \equiv \frac{1}{\sqrt{6}} \sum_{j=0}^{2} e^{i\frac{2\pi}{3}j}(\|\mathbf{G}_j\rangle + \|\bar{\mathbf{G}}_j\rangle)$, | $\|\bar{\psi}_{+,G}\rangle \equiv \frac{1}{\sqrt{6}} \sum_{j=0}^{2} e^{i\frac{2\pi}{3}j}(\|\mathbf{G}_j\rangle - \|\bar{\mathbf{G}}_j\rangle)$, |
| −1 | $\|\psi_{-,G}\rangle \equiv \frac{1}{\sqrt{6}} \sum_{j=0}^{2} e^{-i\frac{2\pi}{3}j}(\|\mathbf{G}_j\rangle + \|\bar{\mathbf{G}}_j\rangle)$, | $\|\bar{\psi}_{-,G}\rangle \equiv \frac{1}{\sqrt{6}} \sum_{j=0}^{2} e^{-i\frac{2\pi}{3}j}(\|\mathbf{G}_j\rangle - \|\bar{\mathbf{G}}_j\rangle)$. |

## III. Results for monolayer transition metal dichalcogenides on twisted bilayer graphene

To gain some quantitative results, it is instructive to apply the above theoretical model to a realistic system and compare it to experimental results. In a TBG moiré pattern, doped carriers will accumulate at **AA**-stacked regions, resulting in a periodic charge distribution [40,43]. We approximate the charge at each **AA** site by a Gaussian wavepacket (see Fig. 1(a,b)), and write the total charge density distribution in the TBG moiré pattern as

$$\rho(\mathbf{R}) = \frac{\nu}{\pi \delta R^2} \sum_l e^{-\frac{(\mathbf{R}-\mathbf{R}_l)^2}{\delta R^2}}. \tag{6}$$

Here $\mathbf{R}_l$ is the center position of $l$-th **AA** site, and the wavepacket width $\delta R$ should vary with the carrier density quantified by $\nu$ which is the filling factor in each moiré supercell. $\nu = 1$ corresponds to a density $2\lambda^{-2}/\sqrt{3} \approx 2.09 \times 10^{11}$ cm$^{-2}$ for $\lambda = 23.5$ nm (the case of $\theta = 0.6°$ in Ref. [40]), and $\nu = 4$ fully fills the first moiré mini-band. $\rho(\mathbf{R})$ in TBG can generate a periodic electrostatic potential $U(\mathbf{R}) = \sum_{\mathbf{G}} e^{i\mathbf{G}\cdot\mathbf{R}} U(G)$ in the adjacent TMDs monolayer, with [44]

$$U(G) \approx \nu \frac{b}{\epsilon \lambda G} \frac{e^{-G^2 \delta R^2/4 - Gd}}{(1+r_0 G)(1+r_0' G) - r_0 r_0' G^2 e^{-2Gd}}. \tag{7}$$

Here $b \equiv \frac{4\pi}{\sqrt{3}\lambda}$ is the length of the primitive reciprocal lattice vector and $d \approx 0.6$ nm is the interlayer spacing between the monolayer TMDs and TBG. $r_0$ and $r_0'$ are screening lengths of monolayer TMDs and TBG, respectively. $r_0 \approx 4.5/\epsilon$ can be obtained from first-principles calculations [45], but $r_0'$ is not known. Considering that the presence of the adjacent TBG can significantly reduce the exciton binding energy in TMDs [37,40], we expect $r_0'$ to be much larger than $r_0$. The wavepacket width $\delta R$ is expected to increase with $\nu$ due to the Pauli exclusion and Coulomb repulsion between carriers, which for simplicity can be approximated by a linear relation. Meanwhile $\delta R$ should be nearly independent on the moiré wavelength for $\lambda \sim 10$ nm or larger, due to the lattice reconstruction effect which keeps the area of **AA** region in TBG fixed ($\sim 2.6$ nm) [40]. Below we set $r_0' = 70/\epsilon$ and $\delta R = 2(1+0.1\nu)$ nm, which are found to give good agreement between theoretical and experimental results.

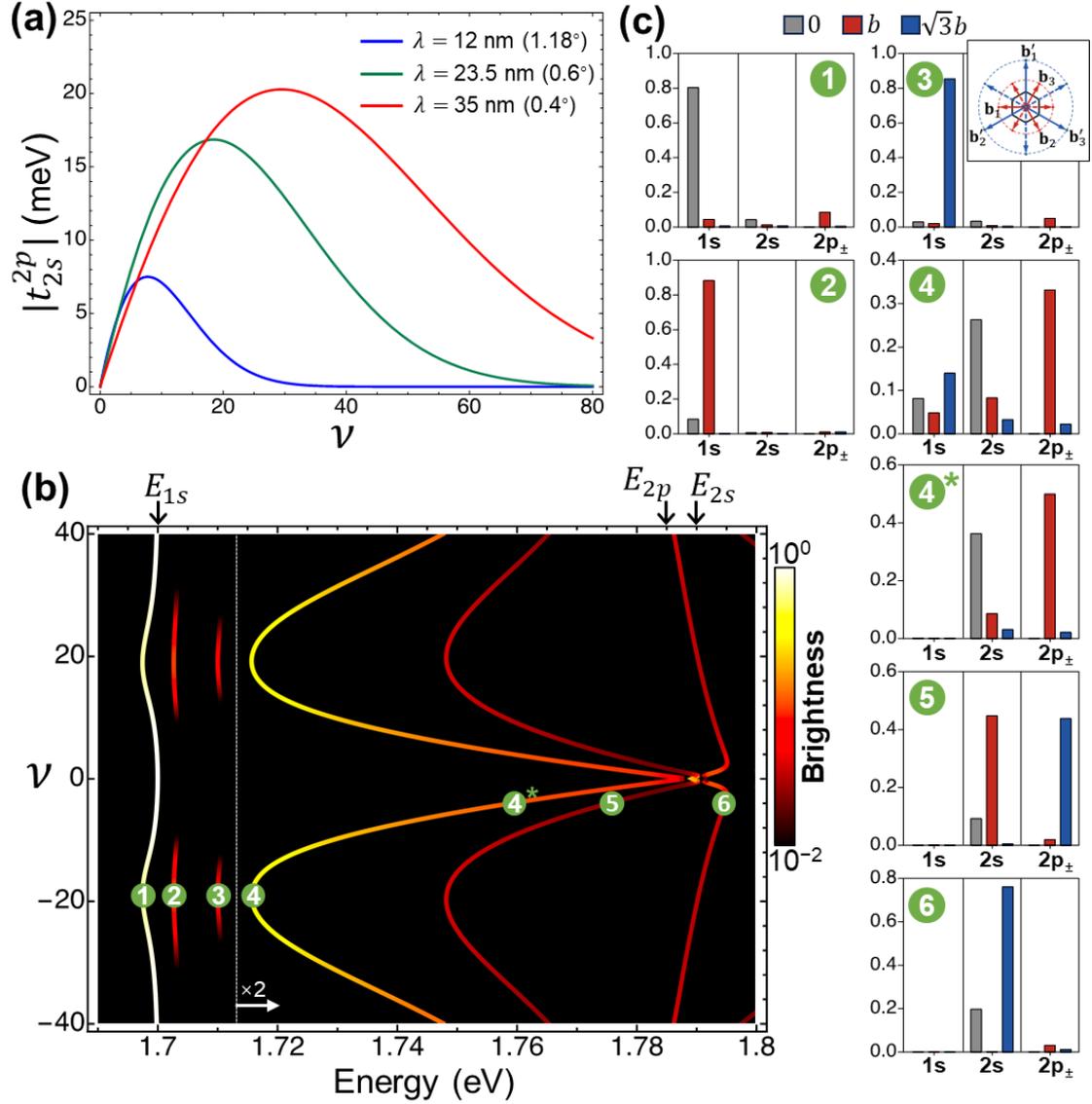

Fig. 2 (a) The coupling strength $|t_{2s}^{2p}|$ as a function of $\nu$ under $\epsilon = 5$ and $\lambda = 12, 23.5$ and 35 nm, which corresponds to 1.18°, 0.6° and 0.4° TBG, respectively. (b) The calculated optical absorption spectra as functions of $\nu$, for $\lambda = 23.5$ nm. (c) The Rydberg state and CoM momentum compositions for the bright states ❶ to ❻ in (b). Different colors denote different magnitudes of the CoM momentum.

The carrier density $\nu$ in TBG can affect the Hamiltonian in Eq. (4). First and most importantly, the resultant electrostatic potential gives rise to finite coupling strengths

$|t_{ns}^{2p}| = U_0 b |\langle 2p_\pm | \hat{\mathbf{r}} | ns \rangle|$ in Eq. (4). $|\langle 2p_\pm | \hat{\mathbf{r}} | ns \rangle|$, determined by the wave function extension of $|ns\rangle$ and $|2p_\pm\rangle$, is expected to be in the order of 1-10 nm. By setting $\epsilon = 5$ (the dielectric constant of the surrounding thick hBN layers in experiments), $|\langle 2p_\pm | \hat{\mathbf{r}} | 2s \rangle| = 7.5$ nm and $|\langle 2p_\pm | \hat{\mathbf{r}} | 1s \rangle| = 1$ nm, values of $|t_{2s}^{2p}| = 7.5 |t_{1s}^{2p}|$ as functions of $v$ under $\lambda = 12, 23.5$ and $35$ nm are shown in Fig. 2(a). We can see that $|t_{2s}^{2p}|$ increases linearly with $v$ in the low-density limit, but decays exponentially under a high density. Second, the screening effect of doped carriers in TBG can correct both the electronic band gap and the exciton binding energy in monolayer TMDs. Experiments have been shown that, the two corrections almost cancel each other for the 1s exciton such that its energy remains constant under different carrier densities with $E_{1s} \approx 1.70$ eV [37]. On the other hand, the 2s exciton red shifts significantly with the increase of the carrier density ($\sim 5$ meV under $10^{12}$ cm$^{-2}$) due to the reduced energy splitting $E_{2s} - E_{1s}$. By fitting the experimental result in Ref. [40], we write $E_{2s} \approx 1.79$ eV $- \sqrt{\frac{2v}{\sqrt{3}\lambda^2}} \times 10^{-12}$ cm$^2 \times 5$ meV.

We emphasize that the purpose of selecting the above parameter values is to reproduce the experimental observations in Ref. [40], and the resultant optical spectrum of excitons from our model under $\lambda = 23.5$ nm is shown in Fig. 2(b) (also see Fig. A1 for a comparison between the theoretical and experimental results). Here the brightness of the eigenstate $|\Phi_{n_X, \mathbf{Q}=0}\rangle$ is characterized by its normalized oscillator strength given by

$$\frac{|\langle \text{vac} | \hat{\mathbf{r}} | \Phi_{n_X, \mathbf{Q}=0} \rangle|^2}{|\langle \text{vac} | \hat{\mathbf{r}} | 1s \rangle|^2} = \left| \langle \mathbf{Q}=0, 1s | \Phi_{n_X, \mathbf{Q}=0} \rangle + \frac{\langle \text{vac} | \hat{\mathbf{r}} | 2s \rangle}{\langle \text{vac} | \hat{\mathbf{r}} | 1s \rangle} \langle \mathbf{Q}=0, 2s | \Phi_{n_X, \mathbf{Q}=0} \rangle \right|^2. \quad (8)$$

Here $\langle \text{vac} | \hat{\mathbf{r}} | ns \rangle$ is the optical dipole with $|\text{vac}\rangle$ denoting the vacuum state, and we set $|\langle \text{vac} | \hat{\mathbf{r}} | 1s \rangle|^2 = 9 |\langle \text{vac} | \hat{\mathbf{r}} | 2s \rangle|^2$ which comes from the numerically calculated 1s and 2s exciton wave functions [46]. We can see that multiple (at least 6) bright branches emerge under the effect of the periodic electrostatic potential. The Rydberg orbital and CoM momentum compositions for the selected states ❶ to ❻ are indicated in Fig. 2(c). The lowest-energy branch with the largest brightness is dominated by $|\mathbf{Q}=0, 1s\rangle$, whose energy exhibits a relatively small shift due to the large difference $E_{2p_\pm} - E_{1s} \approx 90$ meV. The 2nd (3rd) lowest-energy branch with an energy $\approx \frac{\hbar^2 b^2}{2M} + E_{1s}$ ($\approx \frac{3\hbar^2 b^2}{2M} + E_{1s}$) is dominated by $|\psi_{0,b}, 1s\rangle$ ($|\psi_{0,\sqrt{3}b}, 1s\rangle$), which becomes bright due to the finite fraction of $|0, 1s\rangle$ when $v$ falls in a suitable range. The 4th (5th) bright branch is the strong hybridization of $|\mathbf{Q}=0, 2s\rangle$, $|\psi_{0,b}, 2s\rangle$ and $|\bar{\psi}_{\mp,b}, 2p_\pm\rangle$ ($|\mathbf{Q}=0, 2s\rangle$, $|\psi_{0,b}, 2s\rangle$ and $|\bar{\psi}_{\mp,\sqrt{3}b}, 2p_\pm\rangle$), whereas the 6th branch is dominated by $|\psi_{0,\sqrt{3}b}, 2s\rangle$.

When $v$ increases from close to 0, the 4th and 5th branches first red shift linearly, and then blue shift after $v$ reaching a critical value $v_m$. Such a behavior originates from the near degeneracy between $E_{2s}$ and $E_{2p}$ together with the large coupling strength $|t_{2s}^{2p}|$. At a critical carrier density $v_m \approx 20$ which gives the largest $|t_{2s}^{2p}|$ (see the solid green line in Fig. 2(a)), a maximum energy shift $\Delta E_m \sim 70$ meV for the 4th branch is achieved (state ❹ in Fig. 2(b)). These behaviors can well reproduce experimental observations in Ref. [40,43]. Compared to state ❹*, state ❹ exhibits a significant fraction of 1s state due to the small energy difference between ❶ and ❹.

Fig. 3(a) indicates the real-space CoM density distributions $\int d\mathbf{r} |\Phi_{n_X, \mathbf{Q}=0}(\mathbf{R}, \mathbf{r})|^2$ for states ❶ to ❻ in Fig. 2(b), which are found to show standing wave patterns with maximum values located near **AA** sites. It is known that due to the opposite charges of the electron and hole, a spatially homogeneous electrostatic potential only affects the exciton's internal motion but not its CoM motion. The results in Fig. 3(a) indicate that the exciton's CoM motion can be localized by a periodic electrostatic potential that spatially modulates the exciton's internal motion. Considering the strong oscillator strength of $|\mathbf{Q}=0, 1s\rangle$, these bright exciton branches then have nano-patterned spatially inhomogeneous couplings to the optical field. Meanwhile, the coupled internal and CoM motions for exciton state ❹ in Fig. 2(b) can be more directly visualized by the distinct electron-hole relative motions when fixing $\mathbf{R}$ at two different positions. Fig. 3(b) and 3(c) indicate the electron-hole relative wave function for $\mathbf{R}$ located at **AB** and **AA**, respectively. $\Phi_{n_X, \mathbf{Q}=0}(\mathbf{AB}, \mathbf{r})$ is similar to 1s, whereas $\Phi_{n_X, \mathbf{Q}=0}(\mathbf{AA}, \mathbf{r})$ shows the superposition state of 2s and $2p_\pm$.

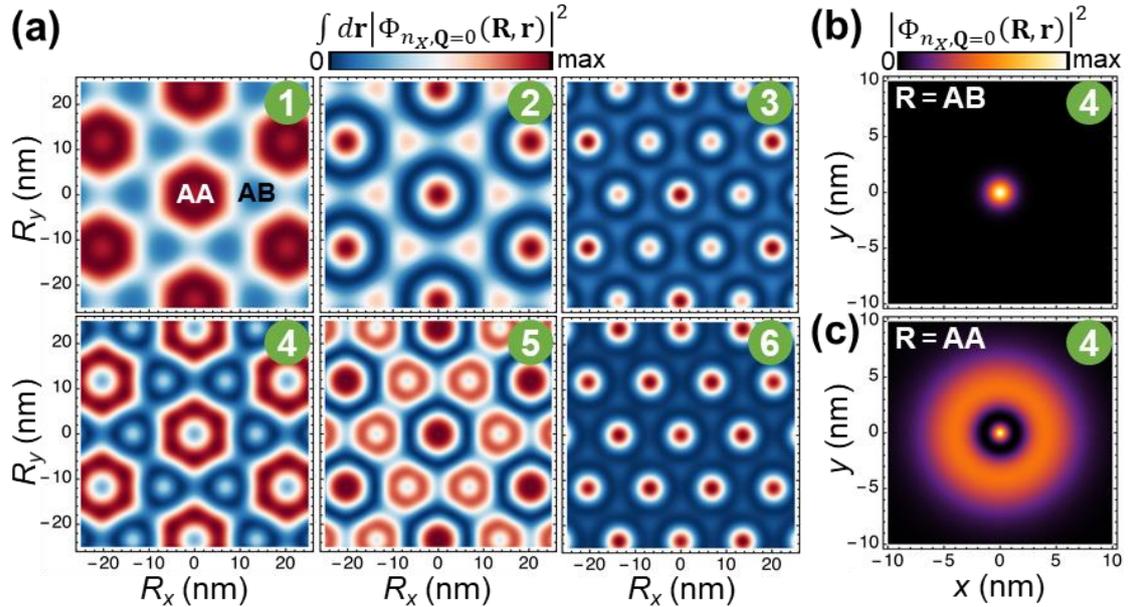

Fig. 3 (a) The real-space CoM density distributions $\int d\mathbf{r} |\Phi_{n_X, \mathbf{Q}=0}(\mathbf{R}, \mathbf{r})|^2$ for the bright exciton states ❶ to ❻ in Fig. 2(b). The density distribution of ❹* is similar to that of ❹. (b) The real-space wave function for the electron-hole relative motion of state ❹ when the CoM position $\mathbf{R}$ is fixed at **AB**. (c) The relative wave function of state ❹ when $\mathbf{R}$ is fixed at **AA**.

We have shown in Fig. 2(a) that the moiré wavelength $\lambda$ can greatly affect the coupling strength $|t_{ns}^{2p}|$. To reveal how the hybrid excitons are influenced by $\lambda$, we calculated the exciton spectra for two other wavelengths distinct from $\lambda = 23.5$ nm. The spectrum under a relatively short wavelength $\lambda = 12$ nm is shown in Figs. 4(a), where the weaker $|t_{ns}^{2p}|$ value results in fewer bright branches compared to Fig. 2(b). Now the two branches with energies $\approx \frac{\hbar^2 b^2}{2M} + E_{1s}$ and $\approx \frac{3\hbar^2 b^2}{2M} + E_{1s}$ become too dark to observe. Meanwhile, the maximum energy shift $\Delta E_m$ at the critical density $\nu_m$ is also significantly smaller (~ 30 meV vs. ~ 70 meV under $\lambda = 23.5$ nm). Fig. 4(b) summarizes the wavelength dependences of $\Delta E_m$ and $\nu_m$. With the increase of $\lambda$, both $\Delta E_m$ and $\nu_m$ become larger. $\nu_m$ has an approximately linear dependence on $\lambda$, whereas $\Delta E_m$ increases sub-linearly with $\lambda$, in good agreement with the experimental observation [40,43]. Fig. 4(c) is the spectrum under a large wavelength $\lambda = 35$ nm, where the lowest-energy state ❶ has a much larger red-shift and significantly weaker brightness compared to that in Fig. 2(b). In contrast, states ❷ and ❸ become much brighter. To see more clearly the underlying mechanism, we turn off the coupling $|\mathbf{Q}, 1s\rangle \leftrightarrow |\mathbf{Q}', 2p_\pm\rangle$ by artificially setting $t_{1s}^{2p} = 0$, and show the resultant spectrum as dashed cyan curves in Fig. 4(c). It indicates that the large wavelength of the periodic electrostatic potential induces a strong coupling between $|\mathbf{Q} = 0, 2s\rangle$, $|\psi_{0,b}, 2s\rangle$ and $|\bar{\psi}_{\mp,b}, 2p_\pm\rangle$, such that the energy of their hybridized state can be shifted to below $E_{1s}$. After taking into account the finite $t_{1s}^{2p}$ value, low-energy states ❶ to ❹ become the strong hybridizations of $|\mathbf{Q} = 0, 1s\rangle$, $|\psi_{0,b}, 1s\rangle$, $|\psi_{0,\sqrt{3}b}, 1s\rangle$, $|\bar{\psi}_{\mp,b}, 2p_\pm\rangle$ and $|\mathbf{Q} = 0, 2s\rangle$. Such a behavior is further confirmed by Fig. 4(d) which shows the Rydberg state and CoM momentum compositions for states ❶ to ❹. All Rydberg states $|1s\rangle$, $|2s\rangle$ and $|2p_\pm\rangle$ exhibit significant fractions in the lowest-energy state ❶.

The above theoretical results change quantitatively with values of parameters like $r_0'$, $|\langle 2p_\pm|\hat{\mathbf{r}}|2s\rangle|$, $\delta R$, etc., but the main conclusion that the exciton internal and center-of-mass motions are coupled does not change. A comparison between different sets of parameters is given in Appendix, where we can see parameters of $r_0' = 70/\epsilon$ nm, $|\langle 2p_\pm|\hat{\mathbf{r}}|2s\rangle| = 7.5$ nm, $|\langle 2p_\pm|\hat{\mathbf{r}}|1s\rangle| = 1$ nm and $\delta R = 2(1 + 0.1\nu)$ nm used in Fig. 2(b) result in optical spectrum in good agreement with the experiment.

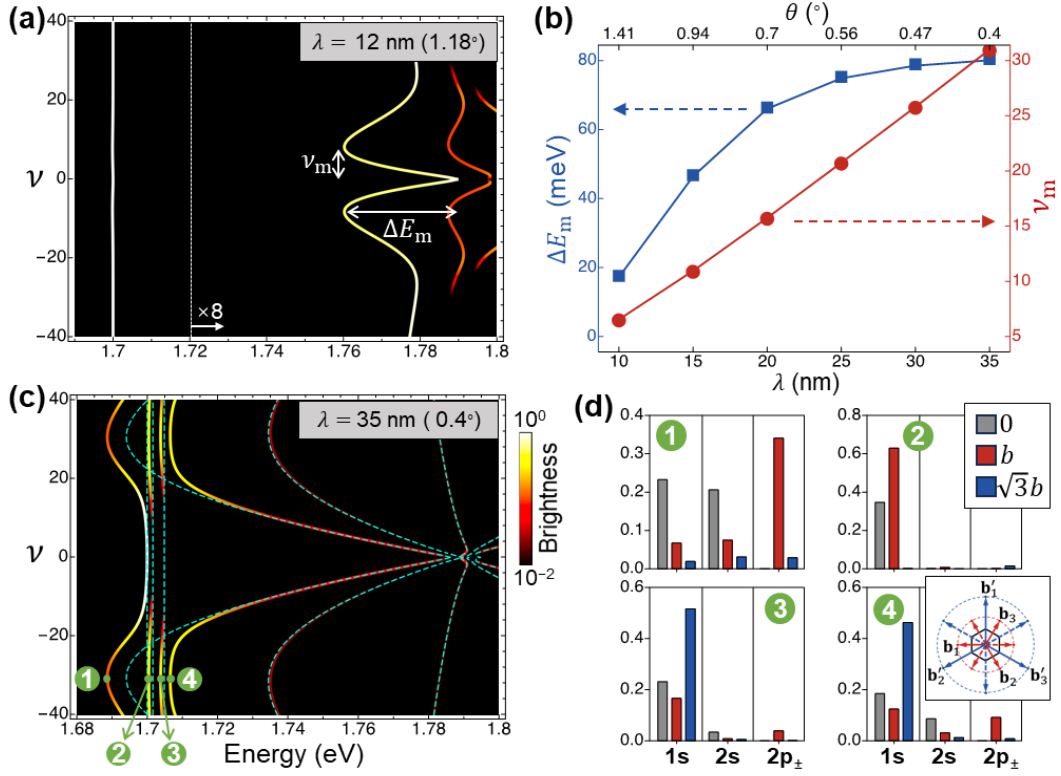

Fig. 4 (a) The calculated optical spectrum under a relatively short wavelength $\lambda$ = 12 nm. (b) The moiré wavelength dependence of the maximum energy shift $\Delta E_m$ and the critical density $\nu_m$. The top horizontal axis indicates the twist angle of TBG for the corresponding $\lambda$ value. (c) The calculated optical spectrum under a large wavelength $\lambda$ = 35 nm. The dashed cyan curves correspond to the result after turning off the coupling $|\mathbf{Q}, 1s\rangle \leftrightarrow |\mathbf{Q}', 2p_\pm\rangle$, i.e., artificially setting $t^{2p}_{1s} = 0$. (d) The Rydberg state and CoM momentum compositions for the four bright states ❶ to ❹ in (c).

## IV. Tunable valley coherence and optical selection rules by an in-plane electric field

In the above analysis, the valley pseudospin of bright excitons in monolayer TMDs is not considered. Below we use $|\mathbf{Q}, nl\rangle|\pm\rangle$ to denote the $\pm K$-valley exciton in $|nl\rangle$ Rydberg state with a CoM momentum $\mathbf{Q}$, then each bright hybrid state shown in Fig. 2(b) and Fig. 4(a,c) exhibits a valley degeneracy. The free bright excitons $|0, ns\rangle|\pm\rangle$ exhibit a circularly polarized valley optical selection rule [13-17], with their coherent superpositions coupling to linearly polarized photons [17]. Meanwhile, two free excitons $|\mathbf{Q}, ns\rangle|\pm\rangle$ in opposite valleys can be coupled by the electron-hole exchange interaction (Fig. 5(a)), with the exchange Hamiltonian given by [1]

$$\hat{H}_{\text{ex}} = \sum_{n,\mathbf{Q}} J_{ns} Q |\mathbf{Q}, ns\rangle\langle \mathbf{Q}, ns|(|+\rangle\langle+| + |-\rangle\langle-|)$$

$$+ \sum_{n,\mathbf{Q}} J_{ns} Q |\mathbf{Q}, ns\rangle\langle \mathbf{Q}, ns|\left(e^{-2i\theta_\mathbf{Q}}|+\rangle\langle-| + e^{2i\theta_\mathbf{Q}}|-\rangle\langle+|\right). \quad (9)$$

Here $\theta_\mathbf{Q}$ is the direction angle of $\mathbf{Q}$, $J_{ns} \propto |\langle \mathbf{r} = 0|ns\rangle|^2$ with $J_{1s} \approx 9 J_{2s} \sim 1$ eV·Å, thus the exchange interaction is significant only for $1s$ state. However, $\hat{H}_{ex}$ doesn't introduce valley coherence to bright excitons with $Q \approx 0$ due to the $\hat{C}_3$-symmetry of monolayer TMDs. For $\hat{C}_3$-symmetric basis states $|\psi_{0,G}, 1s\rangle|\pm\rangle$ whose components have finite CoM momentums (see Table I), their inter-valley couplings from $\hat{H}_{ex}$ also vanish due to the $e^{\pm 2i\theta_\mathbf{Q}}$ phase factor. Thus, the exchange interaction only leads to small energy shifts to bright hybrid excitons but doesn't affect their valley pseudospins. Below we show that when the periodic electrostatic potential introduces a strong coupling between exciton internal and center-of-mass motions, an in-plane electric field can result in non-degenerate exciton eigenstates in the form of valley coherent superpositions, with the valley coherence determined by the field direction.

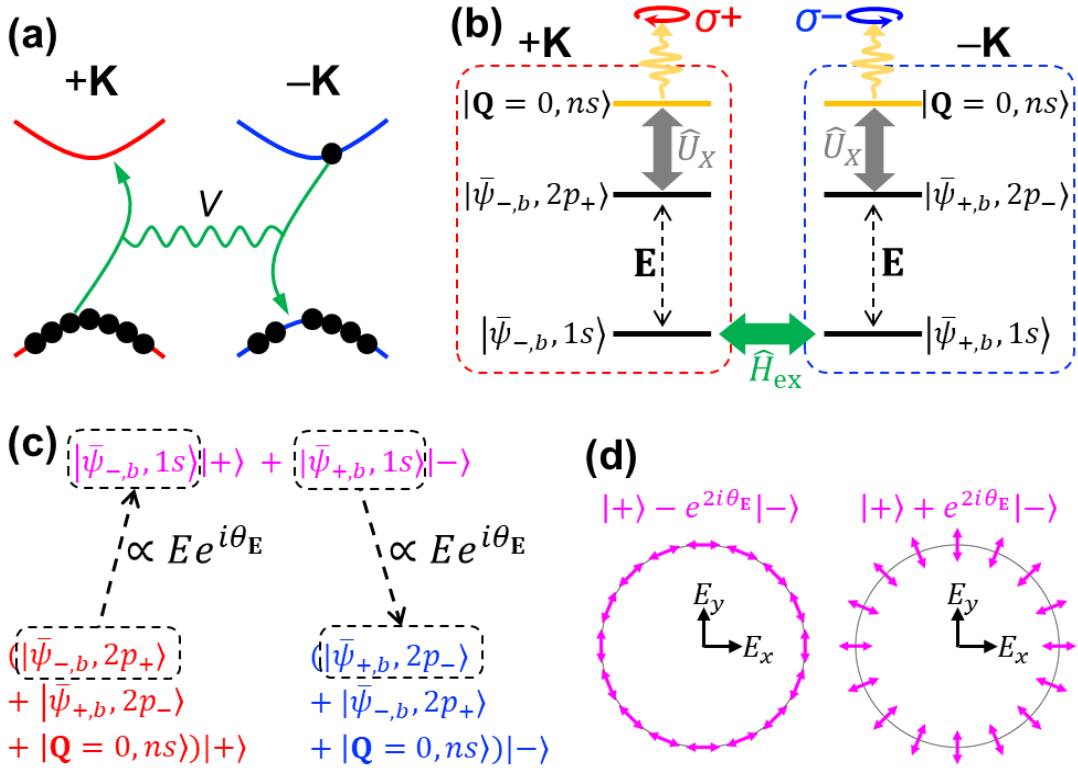

Fig. 5 (a) The diagram for the inter-valley electron-hole exchange interaction. (b) The coupling between $\hat{C}_3$- and inversion-symmetric basis states in $\pm \mathbf{K}$ valleys induced by $\hat{U}_X$, $\hat{H}_{ex}$ and $\mathbf{E}$. $\hat{U}_X$ is the periodic electrostatic potential from the adjacent TBG, $\hat{H}_{ex}$ is the electron-hole exchange interaction, and $\mathbf{E}$ is an externally applied in-plane electric field. $|0, ns\rangle$ in the $\pm \mathbf{K}$ valley can emit a σ± circularly polarized photon. (c) The simplified three-level model. Red (blue) color denotes the state in $+\mathbf{K}$ ($-\mathbf{K}$) valley, whereas purple color denotes a superposition of two valleys. (d) The dependence of emission linear polarizations for the two split valley-coherent excitons with the direction of $\mathbf{E} = (E_x, E_y)$.

$\hat{H}_{ex}$ introduces a finite inter-valley coupling $|\bar{\psi}_{-,b}, 1s\rangle|+\rangle \leftrightarrow |\bar{\psi}_{+,b}, 1s\rangle|-\rangle$ (see Table I for equation forms of $|\bar{\psi}_{\pm,b}\rangle$ ), with the coupling strength

$\langle+|\langle\bar{\psi}_{-,b}, 1s|\hat{H}_{ex}|\bar{\psi}_{+,b}, 1s\rangle|-\rangle = J_{1s}b \approx$ 20 meV. The exchange interaction then results in two valley-coherent superpositions $(|\bar{\psi}_{-,b}, 1s\rangle|+\rangle + |\bar{\psi}_{+,b}, 1s\rangle|-\rangle)/\sqrt{2}$ and $(|\bar{\psi}_{-,b}, 1s\rangle|+\rangle - |\bar{\psi}_{+,b}, 1s\rangle|-\rangle)/\sqrt{2}$ separated by $2J_{1s}b \approx$ 40 meV. Note that $|\bar{\psi}_{-,b}, 1s\rangle$ ($|\bar{\psi}_{+,b}, 1s\rangle$) is $\hat{C}_3$-symmetric with $C_{3X} = -1$ (+1), thus doesn't hybridize to bright excitons with $C_{3X} = 0$ unless the system becomes $\hat{C}_3$-asymmetric.

Applying an in-plane electric field $\mathbf{E} = E(\cos\theta_\mathbf{E}, \sin\theta_\mathbf{E})$ breaks the inversion- and $\hat{C}_3$-symmetries, which couples Rydberg states $|ns\rangle$ and $|2p_\pm\rangle$ in the same valley [47]. This then leads to intra-valley couplings $|\bar{\psi}_{-,b}, 1s\rangle|+\rangle \leftrightarrow |\bar{\psi}_{-,b}, 2p_+\rangle|+\rangle$ and $|\bar{\psi}_{+,b}, 1s\rangle|-\rangle \leftrightarrow |\bar{\psi}_{+,b}, 2p_-\rangle|-\rangle$. When $|\bar{\psi}_{\mp,b}, 2p_\pm\rangle$ becomes strongly hybridized with $|0, ns\rangle$ by the periodic electrostatic potential, $\mathbf{E}$ will indirectly introduce an inter-valley coupling between $|0, ns\rangle|+\rangle$ and $|0, ns\rangle|-\rangle$ (see Fig. 5(b)). This will give rise to valley-coherent bright excitons with linearly polarized optical selection rules. Taking state ❶ in Fig. 4(c) or ❹ in Fig. 2(b) as an example, below we give a rough estimation on the energy splitting between the resultant valley-coherent states. We expect $(|\bar{\psi}_{-,b}, 1s\rangle|+\rangle + |\bar{\psi}_{+,b}, 1s\rangle|-\rangle)/\sqrt{2}$ and $(|\bar{\psi}_{-,b}, 1s\rangle|+\rangle - |\bar{\psi}_{+,b}, 1s\rangle|-\rangle)/\sqrt{2}$ to be less affected by the periodic electrostatic potential, similar to states ❷ in Fig. 2(b) and 4(c). $(|\bar{\psi}_{-,b}, 1s\rangle|+\rangle - |\bar{\psi}_{+,b}, 1s\rangle|-\rangle)/\sqrt{2}$ $((|\bar{\psi}_{-,b}, 1s\rangle|+\rangle + |\bar{\psi}_{+,b}, 1s\rangle|-\rangle)/\sqrt{2})$ has an energy $E_{1s} + \frac{\hbar^2 b^2}{2M}$ ($E_{1s} + \frac{\hbar^2 b^2}{2M} + 2J_{1s}b$) close to ❶ (❹), with a small difference $\Delta \sim$ 10 meV. The six-level model of Fig. 5(b) can then be simplified to a three-level model with close energies shown in Fig. 5(c). For general values of $\Delta$, the resultant perturbative inter-valley coupling strength is $-\frac{\alpha}{8}|\langle 2p_\pm|\hat{\mathbf{r}}|1s\rangle|^2 E^2 e^{-2i\theta_\mathbf{E}}/\Delta$, with $\alpha \sim 0.17$ the fraction of $|\bar{\psi}_{-,b}, 2p_+\rangle$ or $|\bar{\psi}_{+,b}, 2p_-\rangle$ in state ❶ or ❹ (see Fig. 4(c) and Fig. 2(b)). An in-plane electric field of $E$ = 100 V/μm thus leads to a splitting ~10 meV between the formed bright states with valley pseudospins $(|+\rangle \pm e^{2i\theta_\mathbf{E}}|-\rangle)/\sqrt{2}$. This value exceeds the exciton linewidth thus can be observed in experiments. On the other hand, if $\Delta \approx 0$ which can be tuned by the doping density $\nu$ in TBG, then the splitting becomes $\frac{\sqrt{\alpha}}{2}|\langle 2p_-|\hat{\mathbf{r}}|1s\rangle|E$ (~ 10 meV under $E$ = 50 V/μm). These split valley-coherent states have linearly polarized optical selection rules determined by the direction of the applied in-plane electric field. Their linear polarization directions are longitudinal and transverse to $\mathbf{E}$, respectively, which changes $4\pi$ when $\mathbf{E}$ rotates a full circle, see Fig. 5(d). Combined with a magnetic or optical field that splits the energies of $\pm\mathbf{K}$ valleys [18-24], any non-degenerate valley pseudospin state for the bright exciton can be realized.

## V. Conclusion

Our work theoretically investigated the coupling between the exciton internal and CoM motions induced by a periodic electrostatic potential. Taking the TMDs/TBG van der Waals structure as an example, we simulated the optical spectrum of bright excitons in monolayer TMDs under a periodic electrostatic potential generated by the charge distribution in TBG. Due to the CoM momentum dependent coupling between exciton $ns$ and $2p_\pm$ Rydberg states, the oscillator strengths of $ns$ excitons are redistributed into a series of hybrid excitons, resulting in the emergence of additional absorption peaks. By varying the doping density and TBG moiré wavelength, energies of these hybrid eigenstates can be shifted by several tens meV, which well reproduces the recent experimental observations. Wave functions of hybrid eigenstates exhibit nano-patterned spatial modulations, with the electron-hole relative motion depending on the exciton CoM position. Lastly, we propose to use an in-plane electric field to tune the valley pseudospin of the hybridized bright exciton. Our study can serve as a theoretical background for the manipulation of optical properties and valley pseudospin of excitons in two-dimensional semiconductors through proximity controls.

**Acknowledgement.** H.Y. acknowledges support by NSFC under grant No. 12274477, and the Department of Science and Technology of Guangdong Province in China (2019QN01X061). Y.X. acknowledges support by the National Key R&D Program of China (2021YFA1401300).

## Appendix. A comparison between experimental and theoretical optical spectra under different parameter sets

Our theoretical model involves a series of system parameters whose values can quantitatively change the calculated exciton spectrum. In these parameters, the exciton energies $E_{ns}$ can be directly obtained from experiments [40,43], the environmental dielectric constant is given by $\epsilon = 5$ which corresponds to that of bulk hBN since the system is usually encapsulated by thick hBN layers, the screening length of TMDs $r_0 \approx 4.5/\epsilon$ is known from first-principles calculations [45]. However, there are uncertainties for values of other parameters including $|\langle 2p_\pm|\hat{\mathbf{r}}|ns\rangle|$, $r_0'$ (the screening length of TBG) and $\delta R$ (the wavepacket width at **AA** sites of TBG). For a more comprehensive perspective, we vary these parameters and show the resultant exciton spectra in Fig. A1(a-e), with the parameter values summarized in Table II. Different sets of parameters lead to quantitatively different spectra, but main features, including the emergence of multiple bright branches and the energy shifts of higher-energy branches with $v$, are qualitatively the same. For a comparison, the experimentally observed reflectance contrast spectrum is presented in Fig. A1(f). We can see that the optical spectrum in Fig. A1(e) (with parameters the same as Fig. 2(b)) is in good agreement with the experiment. We note that the additional branches in Fig. A1(f) could be due to the spatial inhomogeneity of the TBG moiré pattern in the experiment, which leads to spatially varying $\lambda$ values and the coexistence of multiple energies for a given exciton branch.

Table II. A summary of parameter values used in Fig. A1(a-e). Other parameters not given in the table are $\lambda = 23.5$ nm, $\epsilon = 5$, $r_0 = 4.5/\epsilon$, $E_{1s} = 1.7$ eV and $E_{2s} = 1.79$ eV $- \sqrt{\frac{2\nu}{\sqrt{3}\lambda^2} \times 10^{-12} \text{ cm}^2} \times 5$ meV.

| Figure | $|\langle 2p_\pm|\hat{\mathbf{r}}|2s\rangle|$ | $|\langle 2p_\pm|\hat{\mathbf{r}}|1s\rangle|$ | $r_0'$ | $\delta R$ |
|---|---|---|---|---|
| A1(a) | 10 nm | 1 nm | $70/\epsilon$ nm | $2(1 + 0.1\nu)$ nm |
| A1(b) | 7.5 nm | 1 nm | $50/\epsilon$ nm | $2(1 + 0.1\nu)$ nm |
| A1(c) | 7.5 nm | 1 nm | $70/\epsilon$ nm | $3(1 + 0.1\nu)$ nm |
| A1(d) | 7.5 nm | 1 nm | $70/\epsilon$ nm | $2(1 + 0.5\sqrt{\nu})$ nm |
| 2(b) & A1(e) | 7.5 nm | 1 nm | $70/\epsilon$ nm | $2(1 + 0.1\nu)$ nm |

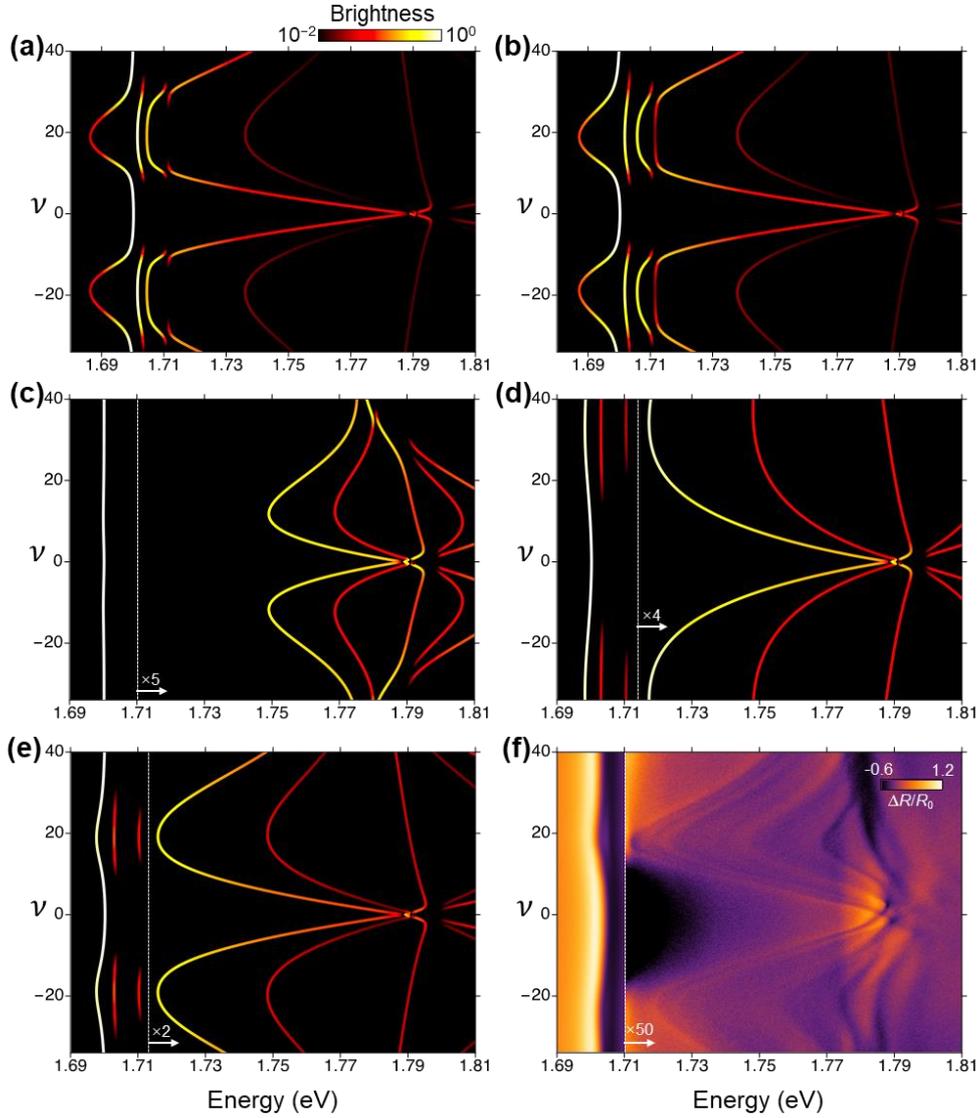

Figure A1. (a-e) Theoretically calculated absorption spectra of excitons under various parameter selections, whose values are summarized in Table II. (f) The experimental reflectance contrast spectrum of a TMDs/TBG sample with $\lambda = 23.5$ nm. Details about the experiment can be found in Ref. [40].

[1] Hongyi Yu, Xiaodong Cui, Xiaodong Xu, and Wang Yao, *Valley excitons in two-dimensional semiconductors*, Natl. Sci. Rev. **2**, 57 (2015).

[2] Gang Wang, Alexey Chernikov, Mikhail M. Glazov, Tony F. Heinz, Xavier Marie, Thierry Amand, and Bernhard Urbaszek, *Colloquium: Excitons in atomically thin transition metal dichalcogenides*, Rev. Mod. Phys. **90**, 021001 (2018).

[3] Nathan P. Wilson, Wang Yao, Jie Shan, and Xiaodong Xu, *Excitons and emergent quantum phenomena in stacked 2D semiconductors*, Nature **599**, 383 (2021).

[4] Alexey Chernikov, Timothy C. Berkelbach, Heather M. Hill, Albert Rigosi, Yilei Li, Ozgur B. Aslan, David R. Reichman, Mark S. Hybertsen, and Tony F. Heinz, *Exciton Binding Energy and Nonhydrogenic Rydberg Series in Monolayer WS2*, Phys. Rev. Lett. **113**, 076802 (2014).

[5] Diana Y. Qiu, Felipe H. da Jornada, and Steven G. Louie, *Optical Spectrum of MoS2: Many-Body Effects and Diversity of Exciton States*, Phys. Rev. Lett. **111**, 216805 (2013).

[6] Ziliang Ye, Ting Cao, Kevin O'Brien, Hanyu Zhu, Xiaobo Yin, Yuan Wang, Steven G. Louie, and Xiang Zhang, *Probing excitonic dark states in single-layer tungsten disulphide*, Nature **513**, 214 (2014).

[7] G. Wang, X. Marie, I. Gerber, T. Amand, D. Lagarde, L. Bouet, M. Vidal, A. Balocchi, and B. Urbaszek, *Giant Enhancement of the Optical Second-Harmonic Emission of WSe2 Monolayers by Laser Excitation at Exciton Resonances*, Phys. Rev. Lett. **114**, 097403 (2015).

[8] Chendong Zhang, Amber Johnson, Chang-Lung Hsu, Lain-Jong Li, and Chih-Kang Shih, *Direct Imaging of Band Profile in Single Layer MoS2 on Graphite: Quasiparticle Energy Gap, Metallic Edge States, and Edge Band Bending*, Nano Lett. **14**, 2443 (2014).

[9] Miguel M. Ugeda *et al.*, *Giant bandgap renormalization and excitonic effects in a monolayer transition metal dichalcogenide semiconductor*, Nat. Mater. **13**, 1091 (2014).

[10] Andreas V. Stier, Kathleen M. McCreary, Berend T. Jonker, Junichiro Kono, and Scott A. Crooker, *Exciton diamagnetic shifts and valley Zeeman effects in monolayer WS2 and MoS2 to 65 Telsa*, Nat. Commun. **7**, 10643 (2016).

[11] Shuo Dong *et al.*, *Direct measurement of key exciton properties: Energy, dynamics, and spatial distribution of the wave function*, Nat. Sci. **1**, e10010 (2021).

[12] Michael K. L. Man *et al.*, *Experimental measurement of the intrinsic excitonic wave function*, Sci. Adv. **7**, eabg0192 (2021).

[13] Kin Fai Mak, Keliang He, Jie Shan, and Tony F. Heinz, *Control of valley polarization in monolayer MoS2 by optical helicity*, Nat. Nanotechnol. **7**, 494 (2012).

[14] Hualing Zeng, Junfeng Dai, Wang Yao, Di Xiao, and Xiaodong Cui, *Valley polarization in MoS2 monolayers by optical pumping*, Nat. Nanotechnol. **7**, 490 (2012).

[15] Ting Cao *et al.*, *Valley-selective circular dichroism of monolayer molybdenum disulphide*, Nat. Commun. **3**, 887 (2012).

[16] G. Sallen *et al.*, *Robust optical emission polarization in MoS2 monolayers through selective valley excitation*, Phys. Rev. B **86**, 081301(R) (2012).

[17] Aaron M. Jones *et al.*, *Optical generation of excitonic valley coherence in monolayer WSe2*, Nat. Nanotechnol. **8**, 634 (2013).

[18] Edbert J. Sie, James W. McIver, Yi-Hsien Lee, Liang Fu, Jing Kong, and Nuh Gedik, *Valley-selective optical Stark effect in monolayer WS2*, Nature Mater. **14**, 290 (2014).

[19] Jonghwan Kim, Xiaoping Hong, Chenhao Jin, Su-Fei Shi, Chih-Yuan S. Chang, Ming-Hui Chiu, Lain-Jong Li, and Feng Wang, *Ultrafast Generation of Pseudo-magnetic Field for Valley*

*Excitons in WSe2 Monolayers*, Science **346**, 1205 (2014).

[20] Ziliang Ye, Dezheng Sun, and Tony F. Heinz, *Optical manipulation of valley pseudospin*, Nat. Phys. **13**, 26 (2017).

[21] G. Aivazian *et al.*, *Magnetic control of valley pseudospin in monolayer WSe2*, Nat Phys **11**, 148 (2015).

[22] A. Srivastava, M. Sidler, A. V. Allain, D. S. Lembke, A. Kis, and A. Imamoglu, *Valley Zeeman effect in elementary optical excitations of monolayer WSe*, Nat Phys **11**, 141 (2015).

[23] D. MacNeill, C. Heikes, K. F. Mak, Z. Anderson, A. Kormányos, V. Zólyomi, J. Park, and D. C. Ralph, *Breaking of Valley Degeneracy by Magnetic Field in Monolayer MoSe2*, Phys Rev Lett **114**, 037401 (2015).

[24] G. Wang, X. Marie, B. L. Liu, T. Amand, C. Robert, F. Cadiz, P. Renucci, and B. Urbaszek, *Control of Exciton Valley Coherence in Transition Metal Dichalcogenide Monolayers*, Phys. Rev. Lett. **117**, 187401 (2016).

[25] Kin Fai Mak and Jie Shan, *Semiconductor moiré materials*, Nat. Nanotech. **17**, 686 (2022).

[26] Chendong Zhang, Chih-Piao Chuu, Xibiao Ren, Ming-Yang Li, Lain-Jong Li, Chuanhong Jin, Mei-Yin Chou, and Chih-Kang Shih, *Interlayer couplings, Moiré patterns, and 2D electronic superlattices in MoS2/WSe2 hetero-bilayers*, Sci. Adv. **3**, e1601459 (2017).

[27] Yi Pan *et al.*, *Quantum-Confined Electronic States Arising from the Moiré Pattern of MoS2-WSe2 Heterobilayers*, Nano Lett. **18**, 1849 (2018).

[28] Hongyi Yu, Gui-Bin Liu, Jianju Tang, Xiaodong Xu, and Wang Yao, *Moiré excitons: From programmable quantum emitter arrays to spin-orbit-coupled artificial lattices*, Sci. Adv. **3**, e1701696 (2017).

[29] Fengcheng Wu, Timothy Lovorn, and A. H. MacDonald, *Topological Exciton Bands in Moiré Heterojunctions*, Phys. Rev. Lett. **118**, 147401 (2017).

[30] Yanhao Tang *et al.*, *Simulation of Hubbard model physics in WSe2/WS2 moiré superlattices*, Nature **579**, 353 (2020).

[31] Emma C. Regan *et al.*, *Mott and generalized Wigner crystal states in WSe2/WS2 moiré superlattices*, Nature **579**, 359 (2020).

[32] Lei Wang *et al.*, *Correlated electronic phases in twisted bilayer transition metal dichalcogenides*, Nat. Mater. **19**, 861 (2020).

[33] Zhaodong Chu *et al.*, *Nanoscale Conductivity Imaging of Correlated Electronic States in WSe2/WS2 Moiré Superlattices*, Phys. Rev. Lett. **125**, 186803 (2020).

[34] Yang Xu, Song Liu, Daniel A Rhodes, Kenji Watanabe, Takashi Taniguchi, James Hone, Veit Elser, Kin Fai Mak, and Jie Shan, *Correlated insulating states at fractional fillings of moiré superlattices*, Nature **587**, 214 (2020).

[35] Artur Branny, Santosh Kumar, Raphaël Proux, and Brian D. Gerardot, *Deterministic strain-induced arrays of quantum emitters in a two-dimensional semiconductor*, Nat. Commun. **8**, 15053 (2017).

[36] Carmen Palacios-Berraquero *et al.*, *Large-scale quantum-emitter arrays in atomically thin semiconductors*, Nat. Commun. **8**, 15093 (2017).

[37] Yang Xu *et al.*, *Creation of moiré bands in a monolayer semiconductor by spatially periodic dielectric screening*, Nat. Mater. **20**, 645 (2021).

[38] Pei Zhao, Chengxin Xiao, and Wang Yao, *Universal superlattice potential for 2D materials from twisted interface inside h-BN substrate*, NPJ 2D Mater. Appl. **5**, 38 (2021).

[39] Dong Seob Kim *et al.*, *Electrostatic moiré potential from twisted hexagonal boron nitride layers*, Nat. Mater. **23**, 65 (2024).

[40] Qianying Hu *et al.*, *Observation of Rydberg moiré excitons*, Science **380**, 1367 (2023).

[41] Zuocheng Zhang *et al.*, *Engineering correlated insulators in bilayer graphene with a remote Coulomb superlattice*, Nat. Mater. **23**, 189 (2024).

[42] Jie Gu, Jiacheng Zhu, Patrick Knuppel, Kenji Watanabe, Takashi Taniguchi, Jie Shan, and Kin Fai Mak, *Remote imprinting of moiré lattices*, Nat. Mater. **23**, 219 (2024).

[43] Minhao He *et al.*, *Dynamically tunable moiré exciton Rydberg states in a monolayer semiconductor on twisted bilayer graphene*, Nat. Mater. **23**, 224 (2024).

[44] M. Danovich, D. A. Ruiz-Tijerina, R. J. Hunt, M. Szyniszewski, N. D. Drummond, and V. I. Fal'ko, *Localized interlayer complexes in heterobilayer transition metal dichalcogenides*, Phys. Rev. B **97**, 195452 (2018).

[45] Ilkka Kylänpää and Hannu-Pekka Komsa, *Binding energies of exciton complexes in transition metal dichalcogenide monolayers and effect of dielectric environment*, Phys. Rev. B **92**, 205418 (2015).

[46] Jianju Tang, Songlei Wang, and Hongyi Yu, *Inheritance of the exciton geometric structure from bloch electrons in two- dimensional layered semiconductors*, Front. Phys. **19**, 43210 (2024).

[47] Bairen Zhu, Ke Xiao, Siyuan Yang, Kenji Watanabe, Takashi Taniguchi, and Xiaodong Cui, *In-Plane Electric-Field-Induced Orbital Hybridization of Excitonic States in Monolayer WSe2*, Phys. Rev. Lett. **131**, 036901 (2023).